# Ferromagnetic Clusters in the Brownmillerite Bilayered Compounds $Ca_{2.5-x}La_xSr_{0.5}GaMn_2O_8$: An Approach to Achieve Layered Spintronics Materials

A. K. Bera [a] and S. M. Yusuf

*Solid State Physics Division, Bhabha Atomic Research Centre, Mumbai 400 085, India*

**ABSTRACT**

We report the effect of La-substitution on the magnetic and magnetotransport properties of Brownmillerite-like bilayered compounds $Ca_{2.5-x}La_xSr_{0.5}GaMn_2O_8$ ($x$ = 0, 0.05, 0.075, and 0.1) by using dc-magnetization, resistivity and magnetoresistance techniques. The Rietveld analysis of the room temperature x-ray diffraction patterns confirms no observable change of average crystal structure with the La-substitution. Both magnetic and magnetotransport properties are found to be very sensitive to the La-substitution. Interestingly, the La-substituted compounds show ferromagnetic-like behavior (due to the occurrence of a double exchange mechanism) whereas, the parent compound is an antiferromagnet ($T_N$~150 K). All compounds show an insulating behavior, in the measured temperature range of 100 – 300 K, with an overall decrease in the resistivity with the substitution. A higher value of magnetoresistance has been successfully achieved by the La-substitution. We have proposed an electronic phase separation model, considering the formation of ferromagnetic clusters in the antiferromagnetic matrix, to interpret the observed magnetization and magnetotransport results for the La-substituted samples. The present study demonstrates an approach to achieve new functional materials, based on naturally occurring layered system like $Ca_{2.5-x}La_xSr_{0.5}GaMn_2O_8$, for possible spintronics applications.

[a] Email: akbera@barc.gov.in

## I. INTRODUCTION

Low dimensional magnetic systems have received a considerable attention in recent years both from the theoretical and experimental points of view as well as for applications in the modern technology (*e.g.*, in the field of high density magnetic recording and spintronics devices, *etc*).[1] The search for new magnetic materials in mixed valence metal oxides that are suitable for possible device applications has raised a number of fascinating fundamental questions concerning spin-, charge-,[2-5] and orbital- ordering.[3, 6, 7]The low dimensional materials with mixed valence metal oxides show a bunch of interesting physical properties such as, colossal magnetoresistance,[8-10] charge- and orbital- ordering, as well as transport properties[11] due to highly anisotropic as well as competitive exchange interactions between the transition metal ions in these systems.[12-15] Moreover, the presence of these behaviours leads to a complex phase diagram (magnetic as well as electronic) for the low dimensional mixed valence metal oxides.

Among the low dimensional systems, the mixed-valance anion-deficient layered perovskite compounds of the type $A_3B'B_2O_8$ [$A$ = La, Ca, Sr, Y and ($B'$, $B$) = alkali or transition metal ions] are of particular interest due to their unusual physical properties and their possible applications in spintronics.[16-20] These Brownmillerite-like compounds crystallize in the orthorhombic crystal structure under the space group $Pcm2_1$. The crystal structure of these compounds consists of two layers (bilayer) of $BO_6$ octahedra separated by a layer of $B'O_4$ tetrahedra along the [010] crystallographic direction. Cations $A$ occupy the positions between these layers in a disordered manner.[16] The magnetic properties of the iso-structural compound $Ca_{2.5}Sr_{0.5}GaMn_2O_8$ show a 3D short-range antiferromagnetic (AFM) ordering above 170 K, a 2D long-range AFM ordering over 160-170 K, and then a 3D long-range AFM ordering below ~ 150 K[18] with an ordered magnetic moment of 3.09 (1) $\mu_B$ per Mn cation at 5 K (aligned along [010]).[16, 17] The presence of a 1:1 ratio of $Mn^{3+}$ and $Mn^{4+}$ ions on the octahedral sites was concluded.[16, 17] It was found that spins are coupled antifferomagnetically in a given *ac*-layer and, within a bilayer, the AFM layers are coupled ferromagnetically along the *b*-axis. Also a local charge ordering was proposed for this system.[17]

The low field magnetoresistance has been observed for systems with an array of artificially grown ferromagnetic-metal / insulator / ferromagnetic-metal junctions due to the spin tunneling through the insulating barier.[21-23] The present system $Ca_{2.5}Sr_{0.5}GaMn_2O_8$ has a typical naturally occurring layered structure with an array of the magnetic $MnO_6$ octahedral bilayers separated by a nonmagnetic $GaO_4$ tetrahedral layer along the crystallographic *b*-direction. However, this system shows an AFM-insulating behavior.[16] Now, if one can make the magnetic layers ($MnO_6$ bilayers) as a ferromagnetic (FM)-metallic in nature, the resulting compound becomes a system with a natural array of the FM-metallic ($MnO_6$ bilayer) / non-magnetic-insulator ($GaO_4$) / FM-metallic ($MnO_6$ bilayer) junctions. It may, then, be possible to tune this type of naturally occurring layered system to show a low field magnetoresistance over a wide temperature range. If so, this type of naturally occurring layered system becomes a potential candidate for device applications as low field sensors, read heads, data storage devices *etc*. For the present system $Ca_{2.5}Sr_{0.5}GaMn_2O_8$, to make the magnetic layers as FM-metallic in nature, one needs to first understand the interaction mechanism between the Mn ions in the $MnO_6$ octahedral bilayers. The electronic configurations of the $Mn^{3+}$ and $Mn^{4+}$ ions in an octahedral environment are $t_{2g}^3e_g^1$; $S = 2$ and $t_{2g}^3e_g^0$; $S = 3/2$, respectively. The two important interactions between Mn ions, namely double exchange (DE)-driven FM interactions and superexchange (SE)-driven AFM interactions, can be possible individually or simultaneously depending upon the ratio between the Mn ions in a given system. In the DE mechanism, [24, 25] a hopping of an itinerant $e_g$ electron occurs between two partially filled *d* orbital of neighboring $Mn^{3+}$ and $Mn^{4+}$ ions, via the orbital overlap $e_g$(Mn) – $2p_\sigma$(O) – $e_g$(Mn), obeying the on-site Hund's rule with the core $t_{2g}$ spins and results a FM-metallic ground state. On the other hand, in the SE mechanism, the spin coupling takes place between Mn ions via $t_{2g}$(Mn) – $2p_\pi$(O) – $t_{2g}$(Mn) orbital overlapping, without any charge transfer. The SE interaction results an AFM-insulating ground state. Therefore, a DE mediated FM interaction could be introduced in the present system $Ca_{2.5}Sr_{0.5}GaMn_2O_8$ by changing the ratio between $Mn^{3+}$ and $Mn^{4+}$ ions from unity. And this ratio can be varied by substitution of a suitable trivalent rare earth ion such as $La^{3+}$ at the divalent alkali metal ion ($Ca^{2+}$/$Sr^{2+}$) site in the parent AFM-insulating compound $Ca_{2.5}Sr_{0.5}GaMn_2O_8$.

In this paper, we have, therefore, studied for the first time the magnetic and magnetotransport properties of the La-substituted compounds $Ca_{2.5-x}La_xSr_{0.5}GaMn_2O_8$ ($x$ = 0, 0.05, 0.075, and 0.1). The $La^{3+}$ ion is chosen because the ionic radii of $La^{3+}$ and $Ca^{2+}$ ions are almost equal (1.16 Å and 1.12 Å, respectively) at the eight fold symmetry and hence, the substitution is expected not to introduce any strong structural effect. An introduction of the DE mediated FM interaction is expected due to the availability of hopping electrons ($e_g$) with the substitution of $La^{3+}$. These FM interactions must compete with the coexisting $t_{2g}$ (Mn) – $2p_\pi$ (O) – $t_{2g}$ (Mn) AFM interaction. As a result, a complex magnetic state, deviated from the AFM state, could be expected for the La-substituted compounds. Furthermore, due to the reduced dimensionality of the magnetic ordering in these compounds,[18] the balance between FM DE and AFM SE interactions is more fragile.[26] Therefore, it is expected that a slight change in dopant content can lead to a significantly different magnetic ground state *i.e.*, a drastic change in the bulk magnetic and transport properties. In the present study, we have observed a distinct change in both temperature and field dependent magnetic properties with a small substitution of $La^{3+}$. A FM-like behavior has been observed for the La-substituted samples whereas, the parent compound is AFM in nature. We have, therefore, succeeded to introduce the DE interaction in the pure AFM system $Ca_{2.5}Sr_{0.5}GaMn_2O_8$ by varying the ratio between Mn ions. The experimental results were interpreted on the basis of an electronic phase separation model where, a formation of FM clusters, dominated by the FM DE interaction, inside the AFM matrix (dominated by the SE interaction) in the La-substituted compounds is considered. Significantly, the present study shows that the magnetic and electronic properties of the layered system $Ca_{2.5}Sr_{0.5}GaMn_2O_8$ can be tuned/optimized by appropriate chemical substitution to achieve new spintronic material based on naturally occurring layered system for practical applications.

## II. EXPERIMENTAL

Polycrystalline samples of $Ca_{2.5-x}La_xSr_{0.5}GaMn_2O_8$ ($x$ = 0, 0.05, 0.075, and 0.1) were synthesized by the conventional solid state reaction method. Stoichiometric amounts of $CaCO_3$, $SrCO_3$, $La_2O_3$, $Ga_2O_3$, and $MnO_2$ were initially mixed using an agate mortar pestle and placed in alumina crucibles. The well ground powders were decarbonated at 1000 ˚C for 24 hours with intermediate grindings and then pressed into a pellet form under 50 Kg/cm$^2$ pressure. The palletized mixtures were heated for total 192 hours at 1100 ˚C in air, the reactants being remixed and re-palletized at frequent intervals.

The powder x-ray diffraction (XRD) measurements were performed on the samples at room temperature using a Cu $K_\alpha$ radiation over the scattering angular range $2\theta$ = 10˚ - 90˚ [in the scattering vector Q (= $4\pi\sin\theta/\lambda$) range of 0.71 - 5.77 Å$^{-1}$ ] with an equal $2\theta$ steps of 0.02˚.

The dc-magnetization measurements were carried out using a commercial (Oxford Instruments) Vibrating Sample Magnetometer (VSM) as a function of temperature and magnetic field. The temperature dependent magnetization for all samples was measured under 100 Oe field over the temperature range of $5 \leq T$ (K) $\leq 320$. For the zero-field-cooled magnetization ($M_{ZFC}$) measurements, the samples were first cooled from the room temperature down to 5 K in a zero field. After applying the magnetic field at 5 K, the magnetization was measured in the warming cycle with the field being on. For the field-cooled-magnetization ($M_{FC}$) measurements, the samples were cooled under the same field (as used for the $M_{ZFC}$) down to 5 K and the $M_{FC}$ was measured in the warming cycle under the same applied field. The M vs *H* curves over all four quadrants (including the initial leg) were measured for all samples at 5 K over ±70 kOe.

The temperature dependent resistivity measurements were performed using the standard four-probe method in the temperature range ~100-320 K under zero applied magnetic field. The magnetoresistance as a function of applied magnetic field was measured at 150, 200 and 300 K. A silver paint was used for making of the connections between the copper wires and the rectangular bars of the samples.

## III. RESULTS AND DISCUSSION

### A. Structural study

The Rietveld refined room temperature X-ray diffraction patterns for the samples $Ca_{2.5-x}La_xSr_{0.5}GaMn_2O_8$ with $x$ = 0, 0.05, 0.075, and 0.1 are shown in the Fig. 1. For a comparison purpose all the diffraction patterns are plotted together as a function of magnitude of the scattering vector. The diffraction patterns were analyzed by the Rietveld refinement technique using the FULLPROF program.[27] The refinement confirmed the single phase formation of all samples. The refinement also confirmed that all samples crystallized in the orthorhombic structure under the space group $Pcm2_1$. The compounds show a layered crystal structure (Fig. 2) with the lattice constants $a \approx 5.43$ Å (~ $\sqrt{2}a_p$), $b \approx 11.36$ Å (~ $3a_p$), and $c \approx 5.30$ Å (~ $\sqrt{2}a_p$) where $a_p$ is the unit cell parameter of the primitive perovskite structure. The derived unit cell parameters along with other structural parameters are given in Table I. The derived lattice constant values are in good agreement with the previously reported values for the similar type of compounds.[17, 18] The observed values of the lattice constants and the unit cell volume for different La-concentrations, shown in the Fig. 3, indicate their negligible variation with the increase of La-concentration. For these compounds, 80% of $Ca^{2+}$ ions are located at the position [4*c* site; (*x*, *y*, *z*)] between the $MnO_6$ octahedral and the $GaO_4$ tetrahedral layers. Whereas, remaining $Ca^{2+}$ ions (20% for parent compound) along with entire $Sr^{2+}$ ions are located at the position [2*b* site; (*x*, ½, *z*)] between the two $MnO_6$ octahedral layers in a given bilayer. For all La substituted compounds, the entire $La^{3+}$ ions are situated at the 2*b* (*x*, 1/2, *z*) site i.e., the positions between two $MnO_6$ octahedral layers within a given bilayer (Fig. 2).

**B. dc-magnetization study**

The temperature dependent $M_{ZFC}$ and $M_{FC}$ curves for all samples under an external applied magnetic field of 100 Oe are shown in the Fig. 4. Broad humps in the $M_{ZFC}$ and $M_{FC}$ curves have been observed in the temperature range ~ 125- 300 K and ~ 100-300 K for the parent and the $x = 0.05$ compounds, respectively [Fig. 4(a)]. A similar type of broad hump in the $M_{ZFC}$ and $M_{FC}$ curves for the parent compound was reported over the same temperature range, 125-300 K.[17] This broad hump-like behavior in the magnetization curves was ascribed to the presence of a short-range spin-spin correlation between Mn ions within a bilayer.[16] Further neutron diffraction study confirmed the presence of an AFM short-range spin-spin correlation.[18] The broad hump-like behavior is more prominent for the sample with $x = 0.05$. Upon further cooling, both $M_{ZFC}$ and $M_{FC}$ curves for the parent compound start increasing at ~125 K. On the other hand, for the sample with $x = 0.05$, the $M_{FC}$ curve starts increasing at ~90 K, whereas $M_{ZFC}$ curve starts increasing at ~ 75 K. Similar type of sharp increase in the $M_{ZFC}$ and $M_{FC}$ curves was already reported for the parent compound[16] and a 3D long-range AFM ordering was found at lower temperatures.[18] Interestingly, the nature of the $M_{ZFC}$ and $M_{FC}$ curves for the compounds with higher La-concentration ($x = 0.075$ and 0.1), is drastically different [Fig. 4(b)]. The $M_{ZFC}$ curves for both $x = 0.075$ and 0.1 samples show a broad hump over the entire temperature range of measurement (5-300 K) with a maxima around 70 K. On the other hand, the $M_{FC}$ curves for these two samples increase gradually with decreasing temperature and attains a near saturation below ~ 50 K, indicating a ferromagnetic-like signature. A significantly higher value of the magnetization has been observed for the samples with $x = 0.075$ and 0.1 as compared to the parent and the $x = 0.05$ samples. The observed $M_{ZFC}$ (*T*) and $M_{FC}$ (*T*) behavior for the samples with $x = 0.075$ and 0.1 can be interpreted considering the presence of FM clusters inside an AFM matrix in an electronic phase separation model. Here we would like to mention that due to the higher contribution from the FM phase in the magnetization, the signature of the AFM transition becomes unnoticeable in the ZFC and FC magnetization curves for these compounds. However, the presence of major AFM phase is clearly evident in the low temperature M vs *H* study (shown later in Fig. 5). During the zero-field cool process, the FM clusters freeze into random orientations, determined by a local anisotropy field, and results in the random orientations of the local magnetization of the individual clusters. On the other hand, during field cool process, the FM clusters align along the direction of applied field and leads to a higher FM-type magnetization. The existence of a peak in the $M_{ZFC}$ vs temperature curves is then interpreted in terms of a competition between the local anisotropy field and the applied magnetic field. Similar argument was used to describe the behaviour of $M_{ZFC}$ and $M_{FC}$ curves of a phase separated system $La_{1-x}Sr_xCoO_3$.[28] For the present compounds with layered crystal structure, the formation of the FM clusters is expected within the bilayers (in the *ac*- plane). Therefore, the observed sluggish temperature response of $M_{ZFC}$ and $M_{FC}$ may be due to the two dimensional nature of the FM clusters within the *ac*-plane. It should be noted that according to this electronic phase separation model, the M (*H*) curve should not saturate at moderate magnetic fields due to the presence of major AFM phase (discussed later). The electronic phase separation has been reported for several half doped perovskite manganites[29-33] where, it is postulated that the system electronically phase separates into FM clusters in an AFM matrix due to the competitive FM DE and AFM SE interactions. The FM DE interaction between the $Mn^{3+}$ and $Mn^{4+}$ ions dominates in the FM cluster phase. On the other hand, the AFM SE interaction between the $Mn^{3+}$ - $Mn^{3+}$ or $Mn^{4+}$ - $Mn^{4+}$ ions dominates in the AFM matrix.

The possibility of the electronic phase separation for the present compounds with La substitution can be viewed in the following way. The parent compound $Ca_{2.5}Sr_{0.5}GaMn_2O_8$ contains both $Mn^{3+}$ and $Mn^{4+}$ ions with a ratio 1:1 maintaining the charge neutrality.[16] Due to the charge ordering,[17] the $e_g$ electrons of $Mn^{3+}$ ions are localized and cannot hop to the vacant $e_g$ state of the neighboring $Mn^{4+}$ ions, thus the exchange mechanism is only AFM SE in nature in the parent system. The similar type of localization of the $e_g$ electrons due to the charge ordering has been observed for the perovskite-manganite systems in the region of $Mn^{3+}$ and $Mn^{4+}$ ratio 1:1 which leads to a AFM-insulating ground state.[34-36] Now, with the substitution of the $La^{3+}$ ions, the ratio between the $Mn^{3+}$ and $Mn^{4+}$ ions increases i.e., the availability of the hopping $e_g$ electrons increases. Hence, the possibility of the hopping of the $e_g$ electrons between the half filled $e_g$ state of the $Mn^{3+}$ ion and empty $e_g$ state of the neighboring $Mn^{4+}$ ion increases.

However, there is a competition between the DE interaction and AFM ordering. In the present case, the ratio between $Mn^{3+}$ and $Mn^{4+}$ ions (1:0.78 for $x$ = 0.1, the maximum La-substituted sample) may not be sufficient to overcome the AFM ordering completely. Therefore, the DE interaction between the $Mn^{3+}$ and $Mn^{4+}$ ions occurs locally and as a result; FM clusters are formed in the AFM matrix.

Figure 5(a) depicts the virgin magnetization curves for all four samples at 5 K. The magnetization for the parent system is very small (~ 0.015 $\mu_B$/f.u. at 70 kOe field) as expected for an AFM compound. Also a linear behavior of magnetization with the applied field confirms the AFM ground state. For the sample with $x$ = 0.05, a higher magnetization has been observed. The M vs $H$ curves for the samples with $x$ = 0.075 and 0.1 show a rapid increase in the low field region (< 10 kOe), indicating a FM-like character, and then a linear response at higher fields. The observed linear contribution in the magnetization at higher field region may be due to the presence of AFM and/or small amount of free spins. A higher slope of the linear part has been observed for La-substituted compounds as compared to the pure AFM parent compound. The observed addition slope can, therefore, be considered due to formation of free spins. The observed smaller value of magnetization suggests a weak ferromagnetism which possibily arises due to the FM-clusters as evident in our temperature dependent magnetization study (Fig. 4). We could successfully fit the field dependence of magnetization for the La-substituted compounds using the following equation (shown in Fig 5(a)).

$$M = M_s L\left(\frac{\mu(H+\lambda M)}{K_B T}\right) + \chi H \qquad (1)$$

where, $L(x)$ is the Langevin function = Coth($x$)-1/$x$, $M_s$ is the saturation magnetization, $\mu$ is the magnetic moment of an individual cluster, $\lambda$ is the mean field constant, $\chi H$ is the linear contribution to magnetization due to AFM phase as well as free spins. The least square fitted values of different parameters are given in the Table II. The derived moment ($\mu$) values of individual FM spin clusters are found to be 6.7(3), 102.2(2), 102.4(3) $\mu_B$ for $x$ = 0.05, 0.075 and 0.1 compounds, respectively. The observed smaller values of $\lambda$ for all compounds suggest that the inter-clusters interactions are very weak. The atomic fractions of the free spins, as estimated from the derived $\chi$ values that are responsible for the addition slope in the M vs. $H$ curves at higher fields as compared to the parent AFM compound ($x$ = 0), are ~ 1.9%, 1.7%, and 1.8% for $x$ = 0.1, 0.075, and 0.05 samples, respectively. Free spins with such a small fraction could arise due to structural defects and/or disorders with the substitution. The saturation magnetization $M_s$ for $x = 0.1$, 0.075, and 0.05 compounds are found to be 0.193(2), 0.174(2), and 0.072(4) $\mu_B$/f.u., respectively. The theoretically expected $M_s$ is given by $M_S = \sum ngS$ $\mu_B$/f.u; where, n is the fraction of an ion in a formula unit (f.u.) with spin $S$, g is the Lande $g$ factor, and the summation is over all ions in a f.u. For the sample with $x = 0.1$, the theoretically expected saturation magnetic moment is 7.1 $\mu_B$/f.u (=1.1×2×2+0.9×2×3/2 $\mu_B$/f.u.) for a ferromagnetically ordered $Mn^{3+}$ ($t_{2g}^3 e_g^1$, $S = 2$) and $Mn^{4+}$ ($t_{2g}^3 e_g^0$, $S = 3/2$) spin-only magnetic moments. So, for the x = 0.1 sample, the volume phase fraction of the FM-clusters is about 2.7 %. For $x = 0.075$ and 0.05 samples, the volume phase fraction of FM-clusters phase have also been estimated as 2.5 % and 1.0 %, respectively. The observed M($H$) behavior for the samples with $x = 0.05$, 0.075 and 0.1 can thus be interpreted with the coexisting FM and AFM phases.

Figure 5(b) shows the M vs $H$ curves at 5 K over the field range of ±70 kOe (covering all four quadrants) for all four samples. An enlarged view of the lower field region of M vs $H$ curves for the samples with $x = 0.05$, 0.075 and 0.1 is depicted in the top-left inset of Fig. 5(b) which shows a hysteresis for the samples with $x = 0.075$ and 0.1. The bottom-right inset of Fig. 5(b) shows the dependence of the remanent magnetization ($M_R$) and coercive field ($H_C$) on the La-concentration. Both $M_R$ and $H_C$ are found to increase sharply for $x > 0.05$. A tendency of saturation is evident for $x > 0.075$. In the M($H$) curves, for the samples with $x = 0.075$ and 0.1, a peculiar field-induced behavior has been observed (Fig. 6). In the 1$^{st}$ quadrant, an irreversibility between the magnetization curves with increasing and decreasing of field (1$^{st}$ and 2$^{nd}$ field sweeps) has been observed in the high field region. In the 5$^{th}$ sweep, the magnetization curve crosses the virgin curve (i.e., the 1$^{st}$ sweep curve) at ~ 0.5 kOe and coincides with the 2$^{nd}$ sweep curve instead of the 1$^{st}$ sweep curve. Therefore, the virgin curve makes a loop with the 2$^{nd}$ and 5$^{th}$ sweep curves at higher field region. But, no such type of irreversibility has been observed in the 3$^{rd}$ quadrant. Similar type of field-induced behavior is also reported for manganite perovskite compounds showing competitive magnetic phases.[37, 38] This behavior can be explained as an increase of the FM-cluster phase fraction with the increasing field at the higher field values (> 5 kOe) due to a conversion of some part of the AFM phase into the FM phase. Then, the converted FM-cluster phase retains the FM state during the next field sweeps (2$^{nd}$ to 5$^{th}$) and results a loop between the 1$^{st}$ and next field sweeps in the 1$^{st}$ quadrant. Similar increase of a FM-cluster phase fraction with increasing field has also been reported in literature for several phase separated systems.[39-42] This observation is again consistent with the proposed phase separation model as discussed earlier.

## C. Resistivity and magnetoresistance study

Figure 7 depicts the temperature dependence of the electrical resistivity for all four samples in the temperature range 100-320 K. For all samples, the resistivity increases monotonically with decreasing temperature and reaches a value about $10^4$ Ω-cm at around 100 K, below this temperature the resistance exceeds the measurement limit (~$10^6$ Ω) of our instrument. The temperature dependence of resistivity indicates that the samples are insulating in nature. Battle *et al.*[16] reported a drop in the resistivity value by an order of magnitude at ~125 K for the parent compound. However, we have not observed any such drop in the resistivity value for any of these samples (Fig. 7). No other report on the resistivity study for similar compounds is available in the literature for further comparison. In the present study, an overall decrease in the resistivity (however, retaining the insulating behavior) has been observed with the La-substitution. The observed insulating behavior of the substituted compounds is consistent with the electronic phase separation model. As the FM-metallic regions are separated from each other by the insulating AFM phase, the crystal as a whole is expected to show an insulating behavior.[43-45] Then, the system is expected to remain in the insulating state up to a critical concentration of the FM phase. Above the critical concentration, the FM clusters begin to make contacts with each others (percolation of the FM state[46]) and it results a metallic state as a whole. The FM phase fraction is observed to be only ~ 3 volume % (estimated from the magnetization study) for the sample with maximum La-concentration ($x$ = 0.1). This FM phase fraction is well below the required percolation threshold concentration (~ 15 % for a 3D system[47]) for a metallic state. Therefore, the insulator to metallic transition is not expected in the present La-substituted compounds.

Now, we discuss below the possible electrical conduction mechanism in these phase-separated compounds. There are several attempts to explain the temperature dependence of electrical resistivity in manganite systems in terms of phase separation model.[48-50] We would like to mention again that these systems have been considered as a mixture of FM-metallic and AFM/paramagnetic-nonmetallic phases. The electrical conduction in these systems is more complicated due to several competitive interactions involved in these systems such as DE interaction, strong electron-lattice coupling, and lattice (polaron) distortion etc. One way to define the electrical conduction in the mixed phase system is by considering simple summation of the relative contributions from all phases. However, this concept is only valid when the different phases are arranged in alternate flat layers, perpendicular to the direction of current flow. For a phase separated system with random distribution of the phases, a phenomenological model by considering an effective medium approximation[51] can be used to explain the electrical conduction where the total resistivity of the phase separated system is supposed to be due to both band electrons and polarons. This model initially used to explain the electrical conduction in the composite materials. Later on this model has also been considered to describe the electrical conduction in disorder materials in the classical limits. The effective resistivity for a typical three dimensional phase separated system (random mixture of metallic and non-metallic phases) is given by the following expression[51-53]

$$\rho = \frac{4\rho_1\rho_2}{\left[(3f-1)\rho_1+(2-3f)\rho_2\right]+\sqrt{\left[(3f-1)\rho_1+(2-3f)\rho_2\right]^2+8\rho_1\rho_2}} \tag{2}$$

where $\rho_1$ and $\rho_2$ are the resistivity of nonmetallic and metallic phase respectively, $f$ is the metallic (FM) phase fraction of the material. To model the experimental data, it is necessary to know the exact analytical expressions of $\rho_1$ and $\rho_2$. However, different groups have conceived different models to define the electrical conduction in the nonmetallic phase for the effective medium approximation method. For example, Rao *et al.*[53] have considered a Mott's variable range hopping model whereas, Ju *et al*.[52] have considered a semiconducting model and, Lakhsmi *et al*.[54] have considered a polaron hopping model. We have observed that the polaron hopping model is more appropriate for our present case. Therefore, the conduction via the insulating phase has been assumed to be represented by a polaron hopping model[55-57] as,

$$\rho_1(T) = \rho_0 T \exp\left(\frac{E_a}{k_B T}\right) \tag{3}$$

where $\rho_0$ is pre-exponential constant and $E_a$ is total activation energy required for activation and transportation of carriers. The conduction via metallic region can be described by the following equation as[52, 58]

$$\rho_2(T) = \rho'_0 + \rho'_2 T^2 + \rho'_{4.5} T^{4.5} \tag{4}$$

where $\rho'_0$ is a residual resistivity arises due to grain and domain boundaries. The $\rho'_2T^2$ resistivity term can be attributed to electron-electron scattering within the Fermi-liquid model and the $\rho'_{4.5}T^{4.5}$ resistivity term has been predicted for an electron-magnon scattering. The experimental resistivity data for all the substituted samples have been fitted well with the Eq. (2) over the entire measured temperature range (as shown in Fig. 7). A polaron hopping model (Eq. (3)) alone describes well the resistivity behavior of the parent ($x = 0$) compound. For all samples, the corresponding least square fitted values of the different parameters are given in the Table III. The fitted values of $f$ [the metallic (FM) phase fractions] show that the fractions of FM-metallic phase increase with the La-substitution (~1.6%, 3%, and 3.4% for $x = 0.05$, 0.075, and 0.1 samples, respectively). The nonmetallic–AFM phase remains as the major phase in these substituted samples and the electrical conduction is predominated by polaron hopping (Eq. (3)). Here, we should mention that the values of FM phase fractions obtained from this resistivity study for the different samples match well with the values estimated earlier from the analysis of the field dependent magnetization study using Eq. (1). The resistivity study again supports the proposed phase separation model conceived for magnetization study where FM clusters inside the AFM matrix was considered.

The corresponding magnetoresistances (MR) can be represented as

$$MR = \left[\rho(H) - \rho(0)\right] / \rho(0) \tag{5}$$

where, $\rho(H)$ is the resistivity under applied magnetic field and $\rho(0)$ is the resistivity under zero magnetic field. Figure 8(a) shows the MR as a function of the applied magnetic field ($H$) for all four samples at 150 K. A negative MR has been observed for all samples and that increases with increase of La-concentration. This trend is found even at room temperature [inset of the Fig. 8(a)]. Now we discuss the possible origins of observed MR in these compounds. In the case of double exchange mediated FM-metallic state, the MR arises due to suppression of spin fluctuations with the increase of applied magnetic field as described below.[59-62] In this case, MR is dominant in the vicinity of the transition temperature. In this process, according to the double-exchange theory, the effective electron (hole) transfer between the neighboring sites depends on the relative angle of the local spins. Application of a magnetic field tends to align the local spins, and the forcedly spin-polarized conduction electron suffers less from the scattering by local spins and becomes more itinerant. As a result, a MR effect has been observed. Above the transition temperature and/or in the range of small amount of FM phase, the conduction mechanism is observed mainly due to the hopping of magnetic polarons. Here the spin of a conducting electron induces a local distortion of the spin lattice and moves on surrounded by this spin polarization. An applied field can increase the spin polarization and subsequently a MR effect occurs. However, in the case of phase separated systems, the MR effect is more complicated due to the additional contributions from the intrinsic inhomogeneties.[63-65]The intrinsic inhomogeneities in the phase separated systems arise due to presence of coexisting competing phases (FM and AFM) and play an important role on the MR. A random resistor network method has been used to explain the MR in such phase separated systems.[64, 66] In additional to these effects, another effect has been observed for some polycrystalline samples due to the presence of grain boundaries. In this process, the observed negative MR is due to spin-polarized tunneling through an insulating grain boundary.[59-62] With the DE mechanism, electrons are able to move easily when the spins of the ions ($Mn^{3+}$ and $Mn^{4+}$) are parallel, and cannot move if they are antiparallel. As a consequence, the magnetic disorder in the interface region will sharply increase the resistance of the grain boundaries and forms an insulating barrier through which spin-polarized tunneling happens. The tunneling probability of electrons through the grain boundaries depends on the relative orientation of the magnetization directions of neighboring grains, which can be considerably altered by the application of a magnetic field. This results in a sharp drop of resistance in the low applied fields. Therefore, a low field magnetoresistance occurs. Since in our present case, the polaron hopping activation process plays a major role, it is obvious that the magnetoresistance of these electronic phase separated systems occurs mainly due to the change of the activation energy with the application of an external magnetic field.[67] Therefore, from the Eqs. (3) & (5), we have

$$MR = \exp\left[\frac{E_a(H) - E_a(0)}{K_B T}\right] - 1 \quad (6)$$

Now, we can consider the field dependence of the activation energy in a general way as follows

$$E_a(H) = E_a(0)\left[1 - (\alpha H)^{\beta}\right] \quad (7)$$

where, $\alpha$ is proportionality constant and $\beta$ is an exponent. Then, the Eq. (4) becomes

$$MR = \exp\left[-\tfrac{E_a(0)}{K_B T}(\alpha H)^{\beta}\right] - 1 \quad (8)$$

Figures 8(b) and 8(c) show the field dependency of the MR at 300, 200 and 150 K for the samples with $x$ = 0.075 and 0.1, respectively. The solid lines correspond to the fitting of the observed data with the Eq. (8). The fitted parameters are given in the Table IV. An increase in the $\beta$ values has been observed with decreasing temperature. The observed increase of the $\beta$ value with decreasing temperature for both compounds ($x$ = 0.075 and 0.1) [Insets of the Fig. 8(b) & 8(c), respectively] suggests a stronger field dependency of the activation energy [Eq. (7)]. The observed stronger field dependence of the activation energy with decreasing temperature may be due to the reduction of thermal spin fluctuations at lower temperatures.

In the present study, the aimed FM state has thus been attained successfully in terms of FM-cluster state by the La-substitution. However, due to very small amount (~ 3 volume %) of the FM cluster phase, the low field magnetoresistance is quite low. Nevertheless, we have succeeded to tune the magnetic and electronic properties of the layered system $Ca_{2.5}Sr_{0.5}GaMn_2O_8$ (with bilayers of $MnO_6$ octahedra separated by a $GaO_4$ tetrahedral layer) by $La^{3+}$ substitution at the $Ca^{2+}$-site. In order to achieve a larger FM phase fraction, a higher $La^{3+}$ substitution in $Ca_{2.5}Sr_{0.5}GaMn_2O_8$ is essential. However, in our study we have found secondary phases with higher La concentrations ($x > 0.1$). By taking an extra precaution, it may be possible to prepare these samples with a higher La concentration as the ionic radii of $Ca^{2+}$ (1.12 Å) and $La^{3+}$ (1.1.6 Å) are almost equal. Thus, it is possible to prepare new functional materials, suitable for spintronics applications, based on this type of naturally grown layered materials, by making the alternative layers as FM-metallic and nonmagnetic-insulating.

## IV. SUMMARY AND CONCLUSIONS

In summary, we have prepared single phase polycrystalline samples of $Ca_{2.5-x}La_xSr_{0.5}GaMn_2O_8$ ($x$ = 0, 0.05, 0.075, and 0.1). It is confirmed that La-substitution has no significant effect on the crystal structure. The magnetic and magnetotransport properties are found to be very sensitive to the electron doping ($La^{3+}$ substitution at the divalent $Ca^{2+}$ site) i.e., the ratio between the $Mn^{3+}$ and $Mn^{4+}$ ions. We have succeeded to introduce FM DE interaction in $Ca_{2.5-x}La_xSr_{0.5}GaMn_2O_8$ compounds by the substitution. As a result, a FM-like behavior in magnetization and an enhancement in the MR have been observed with the La-substitution. All compounds are found to be an insulating type; however, an overall decrease in the resistivity value has been obtained. The temperature dependence of the resistivity curves has been explained by using an effective medium approximation method. The observed changes of the magnetic and magnetotransport behaviors such as, the appearance of ferromagnetic clusters and enhancement of magnetoresistance and their field dependencies have been successfully discussed based on an electronic phase separation model. The FM clusters inside the AFM matrix appear due to a competition between the coexisting FM DE interactions and AFM SE interactions in the La-substituted samples. We have shown that the tuning/modification of various physical properties is possible by chemical substitution in this type of naturally grown layered system to achieve suitable functional materials for practical applications. This concept of introduction of FM interaction in these Brownmillerite-like layered systems to attain the alternating “FM-metallic” and non-magnetic-insulating layers, therefore, would be helpful to model/prepare new functional materials based on naturally occurring layered materials for their device applications.

## ACKNOWLEDGMENTS

A. K. B. thanks A. Jain for his valuable suggestions in this work. A. K. B. also acknowledges the help provided by Sher Singh and R. Chitra for preparing the samples and performing the X-ray diffraction experiments, respectively.

**TABLES**

TABLE I. The Rietveld refined unit cell parameters, atomic positions, isotropic thermal parameters, and $\chi^2$ for the samples $Ca_{2.5-x}La_xSr_{0.5}GaMn_2O_8$ with $x$ = 0, 0.05, 0.075, and 0.1.

| | ***x* = 0** | ***x* = 0.05** | ***x* = 0.075** | ***x* = 0.1** |
|---|---|---|---|---|
| *a* | 5.4364 (2) | 5.4359 (3) | 5.4352 (1) | 5.4372 (2) |
| *b* | 11.3624 (5) | 11.3551 (7) | 11.3655 (3) | 11.3608 (5) |
| *c* | 5.3019 (2) | 5.3054 (3) | 5.3021 (1) | 5.3040 (2) |
| **Ca** | | | | |
| (4*c*) | | | | |
| *x/a* | 0.2265 (9) | 0.2225 (10) | 0.2267 (8) | 0.2235 (8) |
| *y/b* | 0.1860 (4) | 0.1877 (5) | 0.1845 (3) | 0.1844 (3) |
| *z/c* | 0.5090 (3) | 0.5066 (9) | 0.4959 (15) | 0.4974 (5) |
| $B_{iso}$ | 0.82 (5) | 0.83 (4) | 0.80 (5) | 0.80 (6) |
| **Ca/La/Sr** | | | | |
| (2*b*) | | | | |
| *x/a* | 0.2416 (9) | 0.2450 (11) | 0.2415 (6) | 0.2402 (7) |
| *y/b* | 0.5 | 0.5 | 0.5 | 0.5 |
| *z/c* | 0.4998 (2) | 0.5019 (8) | 0.4893 (8) | 0.4911 (6) |
| $B_{iso}$ | 0.48 (3) / 0.36 (4) | 0.33 (3) / 0.48 (3) / 0.36 (5) | 0.31 (7) / 0.48 (4) / 0.51 (6) | 0.32 (5) / 0.50 (5) / 0.52 (5) |
| **Ga** | | | | |
| (2*a*) | | | | |
| *x/a* | 0.3185 (9) | 0.3207 (5) | 0.3186 (7) | 0.3169 (8) |
| *y/b* | 0 | 0 | 0 | 0 |
| *z/c* | 0.0386 (2) | 0.0314 (6) | 0.0371 ( 2) | 0.0395 (9) |
| $B_{iso}$ | 0.87 (4) | 0.87 (5) | 0.89 (4) | 0.90 (2) |
| **Mn** | | | | |
| (4*c*) | | | | |
| *x/a* | 0.2566 (8) | 0.2570 (12) | 0.2565 (6) | 0.2546 (6) |
| *y/b* | 0.3291 (3) | 0.3292 (4) | 0.3295 (3) | 0.3285 (2) |
| *z/c* | 0 | 0 | 0 | 0 |
| $B_{iso}$ | 0.25 (1) | 0.25 (4) | 0.20 (2) | 0.20 (4) |
| **O1** | | | | |
| (2*a*) | | | | |
| *x/a* | 0.3519 (5) | 0.3607 (5) | 0.3498 (3) | 0.3679 (3) |
| *y/b* | 0 | 0 | 0 | 0 |
| *z/c* | 0.3955 (5) | 0.3695 (6) | 0.4140 (4) | 0.4121 (4) |
| $B_{iso}$ | 1.29 (3) | 1.39 (2) | 1.26 (5) | 1.31 (4) |
| **O2** | | | | |
| (2*b*) | | | | |
| *x/a* | 0.2956 (3) | 0.2960 (5) | 0.3115 (3) | 0.3115 (3) |
| *y/b* | 0.5 | 0.5 | 0.5 | 0.5 |
| *z/c* | 0.0434 (7) | 0.0394 (9) | 0.0219 (6) | 0.0252 (5) |
| $B_{iso}$ | 1.27 (3) | 1.27 (5) | 1.13 (4) | 1.20 (4) |
| **O3** | | | | |
| (4*c*) | | | | |
| *x/a* | 0.1813 (2) | 0.1790 (3) | 0.1834 (2) | 0.1917 (3) |
| *y/b* | 0.1390 (1) | 0.1465 (9) | 0.1518 (1) | 0.1485 (1) |
| *z/c* | 0.0383 (5) | 0.0341 (6) | 0.0253 (4) | 0.0183 (6) |
| $B_{iso}$ | 0.99 (3) | 0.99 (2) | 0.46 (2) | 0.61 (4) |
| **O4** | | | | |
| (4*c*) | | | | |
| *x/a* | 0.0151 (5) | 0.0201 (4) | 0.0418 (7) | 0.0376 (9) |
| *y/b* | 0.3398 (1) | 0.3467 (2) | 0.3480 (1) | 0.3482 (1) |
| *z/c* | 0.2469 (8) | 0.2447 (9) | 0.2453 (5) | 0.2408 (6) |
| $B_{iso}$ | 0.85 (3) | 0.85 (4) | 0.91 (5) | 0.81 (1) |
| **O5** | | | | |
| (4*c*) | | | | |
| *x/a* | 0.4786 (5) | 0.4656 (4) | 0.4807 (4) | 0.4821 (4) |
| *y/b* | 0.3072 (1) | 0.3096 (7) | 0.3120 (9) | 0.3104 (9) |
| *z/c* | 0.2763 (8) | 0.2295 (7) | 0.2553 (8) | 0.2593 (7) |
| $B_{iso}$ | 0.88 (2) | 0.88 (3) | 0.77 (3) | 0.84 (2) |
| $\chi^2$ | 2.91 | 2.11 | 2.67 | 2.65 |

TABLE II. The fitted parameters $M_S$, $\mu$, $\chi$, and $\lambda$, derived from the magnetization data.

| **Sample** | ***Ms*** **($\mu_B$/f.u.)** | ***$\mu$*** **($\mu_B$)** | **$\chi$ [$\mu_B$ (f.u.)$^{-1}$kOe$^{-1}$]** | **$\lambda$** |
|---|---|---|---|---|
| $x$=0.05 | 0.072(4) | 6.7(3) | 0.316(5) | 0.005(2) |
| $x$=0.075 | 0.174(2) | 102.2(2) | 0.304(3) | 0.051(4) |
| $x$=0.1 | 0.193(2) | 102.4(3) | 0.347(4) | 0.063(2) |

TABLE III. The parameters as deduced from the electrical transport data.

| **Sample** | **$\rho_0$ ($\times 10^{-5}$ Ω cm)** | **$E_a$ (*meV*)** | **$\rho'_0$ ($\times 10^{-4}$ Ω cm)** | **$\rho'_2$ ($\times 10^{-8}$ Ω cm K$^{-2}$)** | **$\rho'_{4.5}$ ($\times 10^{-12}$ Ω cm K$^{-4.5}$)** | ***f*** |
|---|---|---|---|---|---|---|
| $x$=0 | 6.9(2) | 158.8(7) | ----- | ------ | ------ | ----- |
| $x$=0.05 | 4.1(3) | 157.7(9) | 14.5(4) | 11.8(4) | 13.6(6) | 0.016(2) |
| $x$=0.075 | 2.3(5) | 145.5(4) | 7.6(5) | 3.4(3) | 4.4(6) | 0.030(3) |
| $x$=0.1 | 2.0(4) | 132.9(6) | 4.6(4) | 2.2(4) | 3.2(5) | 0.034(3) |

TABLE IV. The fitted $\alpha$ and $\beta$ parameters, derived from the magnetoresistance data.

| **Temperature (K)** | ***x* = 0.075** | | ***x* = 0.1** | |
|---|---|---|---|---|
| | **$\alpha$ (kOe$^{-1}$)** | **$\beta$** | **$\alpha$ (kOe$^{-1}$)** | **$\beta$** |
| 150 | 0.00063(2) | 1.755(4) | 0.00083(1) | 1.878(8) |
| 200 | 0.00045(2) | 1.639(11) | 0.00054(2) | 1.737(8) |
| 300 | 0.00028(2) | 1.509(5) | 0.00037(2) | 1.613(3) |

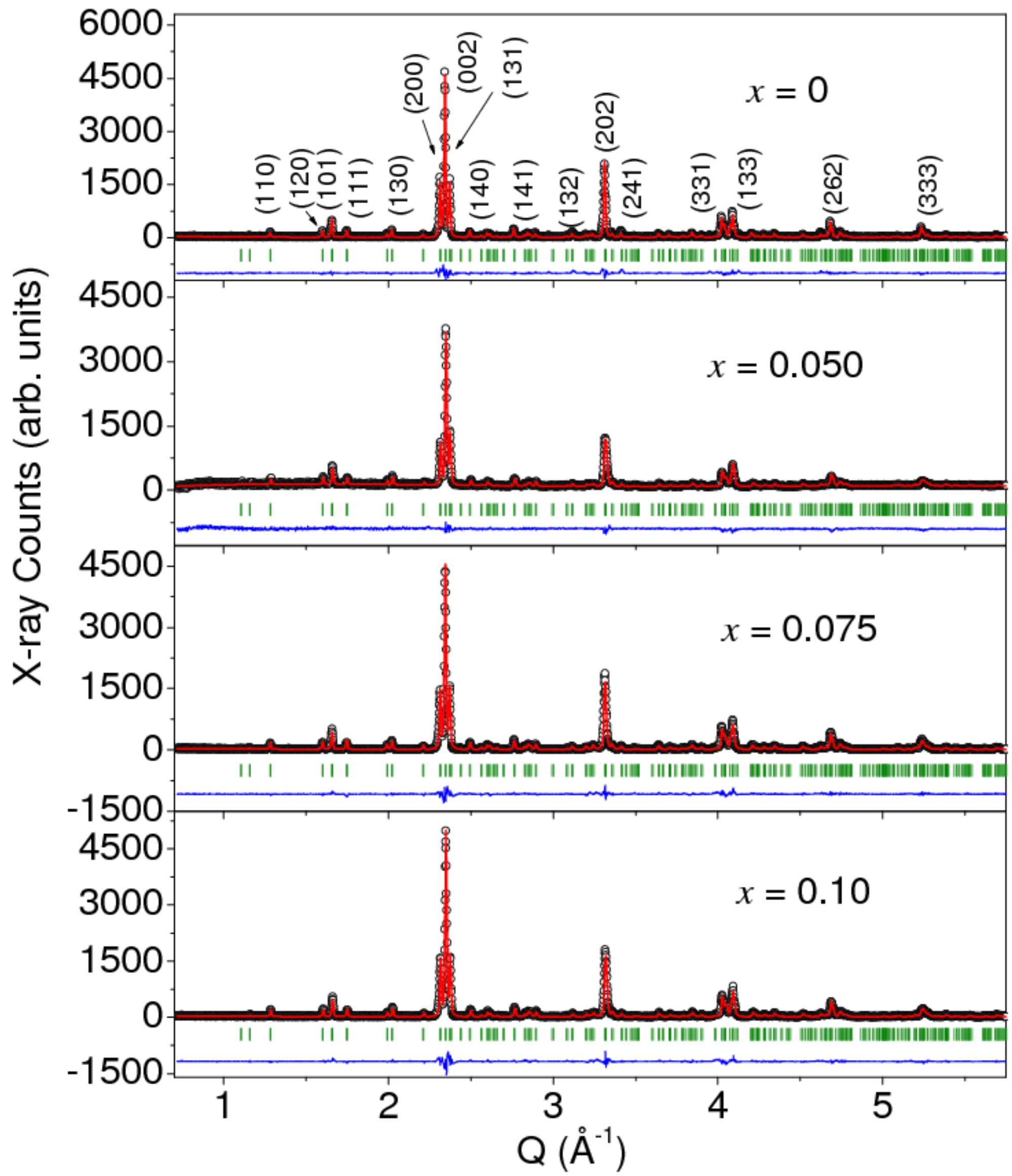

FIG. 1. (Color online) Observed (open circles) and calculated (solid lines) x-ray diffraction patterns of $Ca_{2.5-x}La_xSr_{0.5}GaMn_2O_8$ ($x$ =0, 0.05, 0.075, and 0.10) at room temperature. Solid line at the bottom of each panel shows the difference between observed and calculated patterns. Vertical lines show the position of Bragg peaks. The (*hkl*) values corresponding to intense Bragg peaks are also listed.

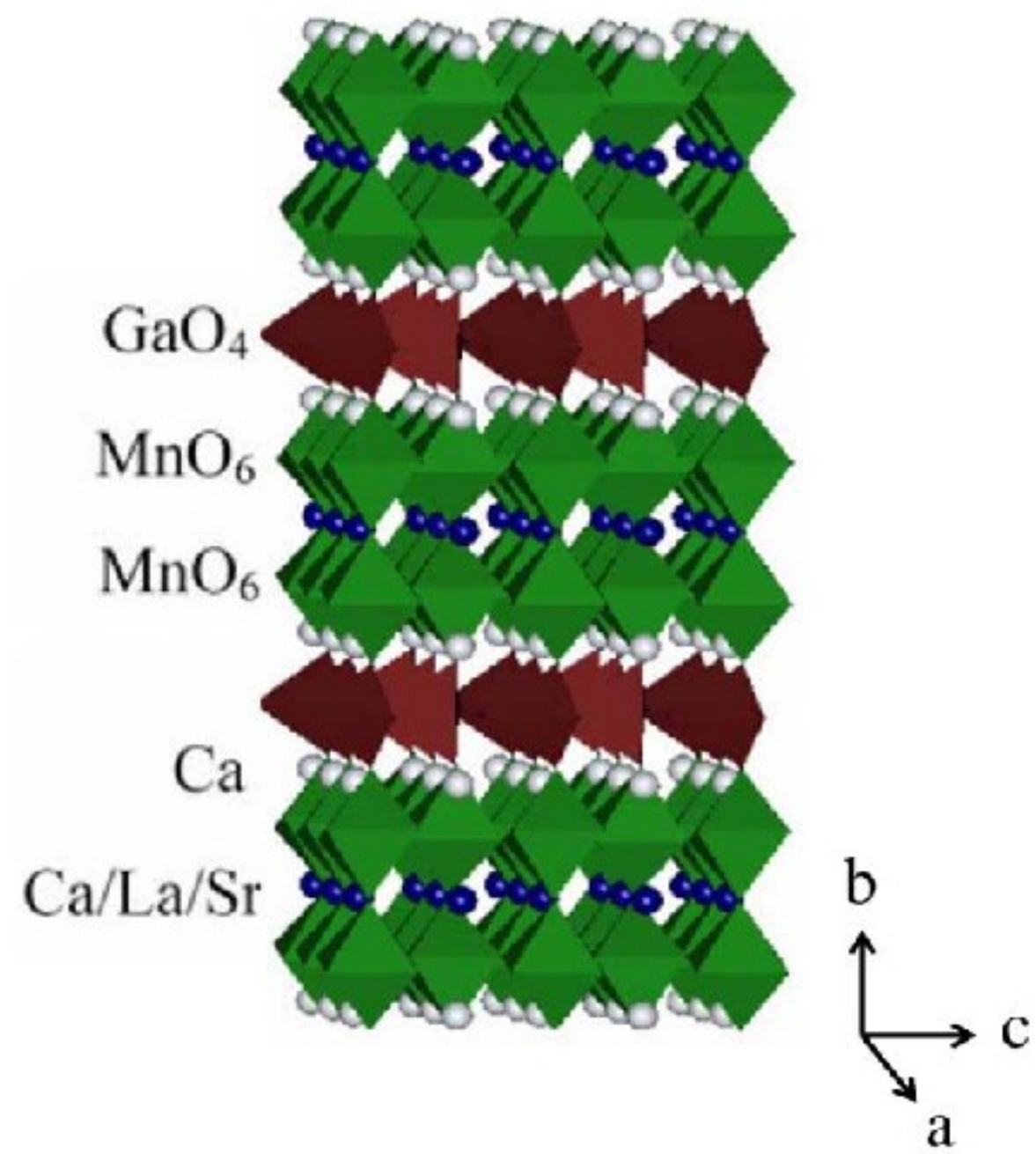

FIG. 2. (Color online) The crystal structure of the compounds $Ca_{2.5-x}La_xSr_{0.5}GaMn_2O_8$.

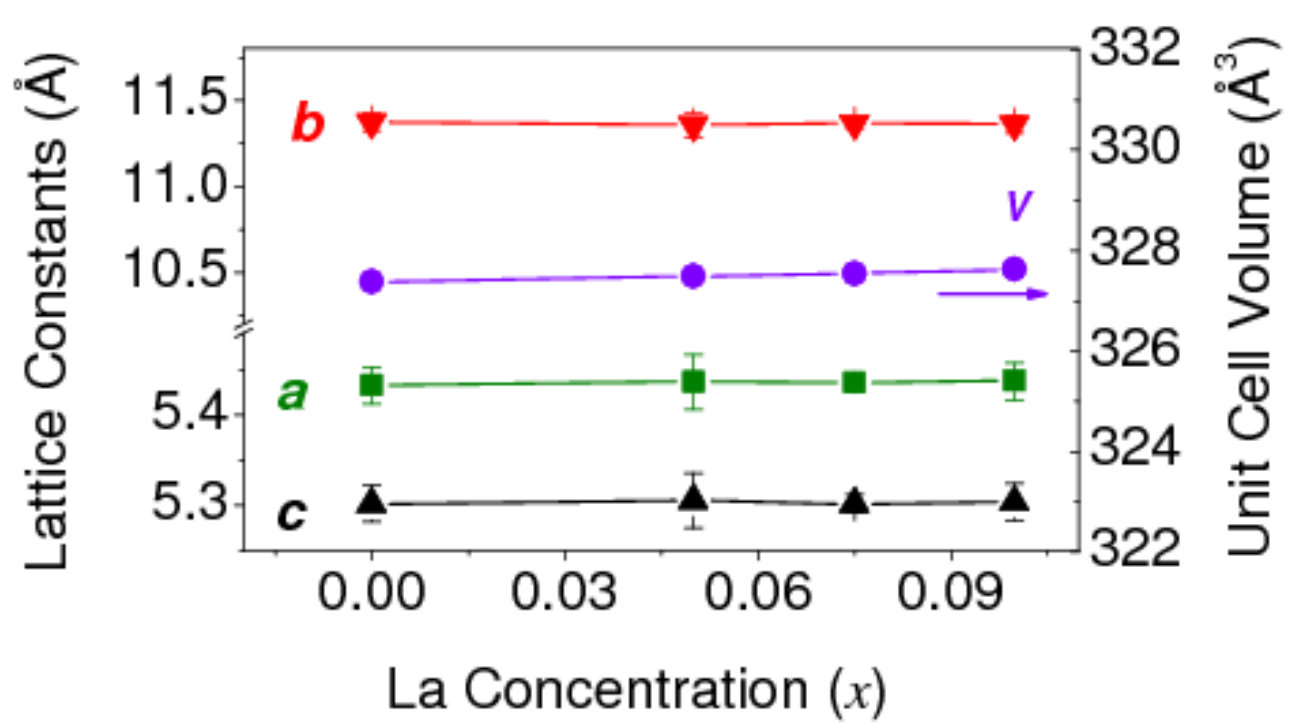

FIG. 3. (Color online) The La concentration dependence of lattice constants and unit cell volume.

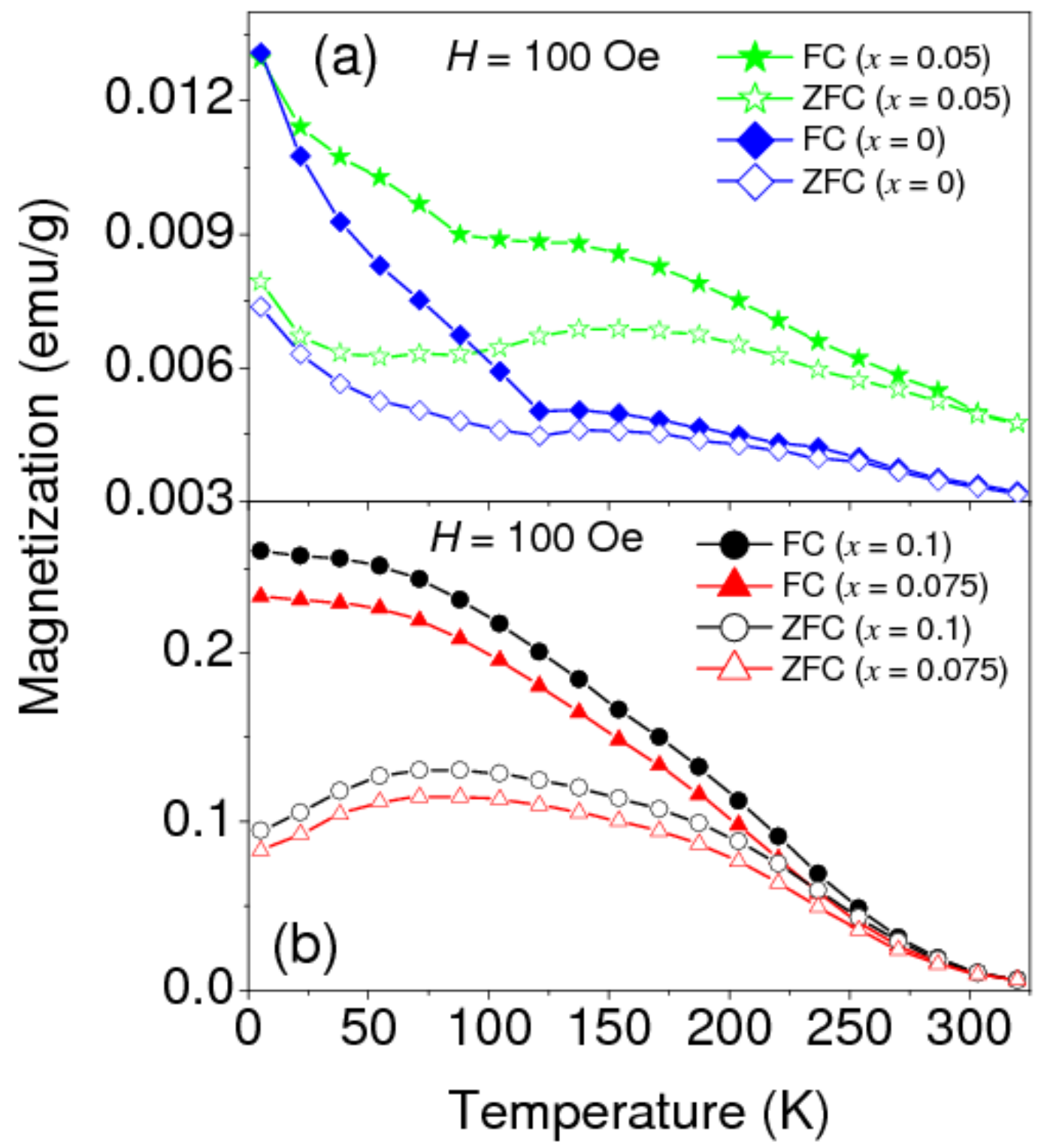

FIG. 4. (Color online) The temperature dependent $M_{ZFC}$ and $M_{FC}$ curves for $Ca_{2.5-x}La_xSr_{0.5}GaMn_2O_8$ ($x$ =0, 0.05, 0.075, and 0.10) compounds under 100 Oe applied magnetic field.

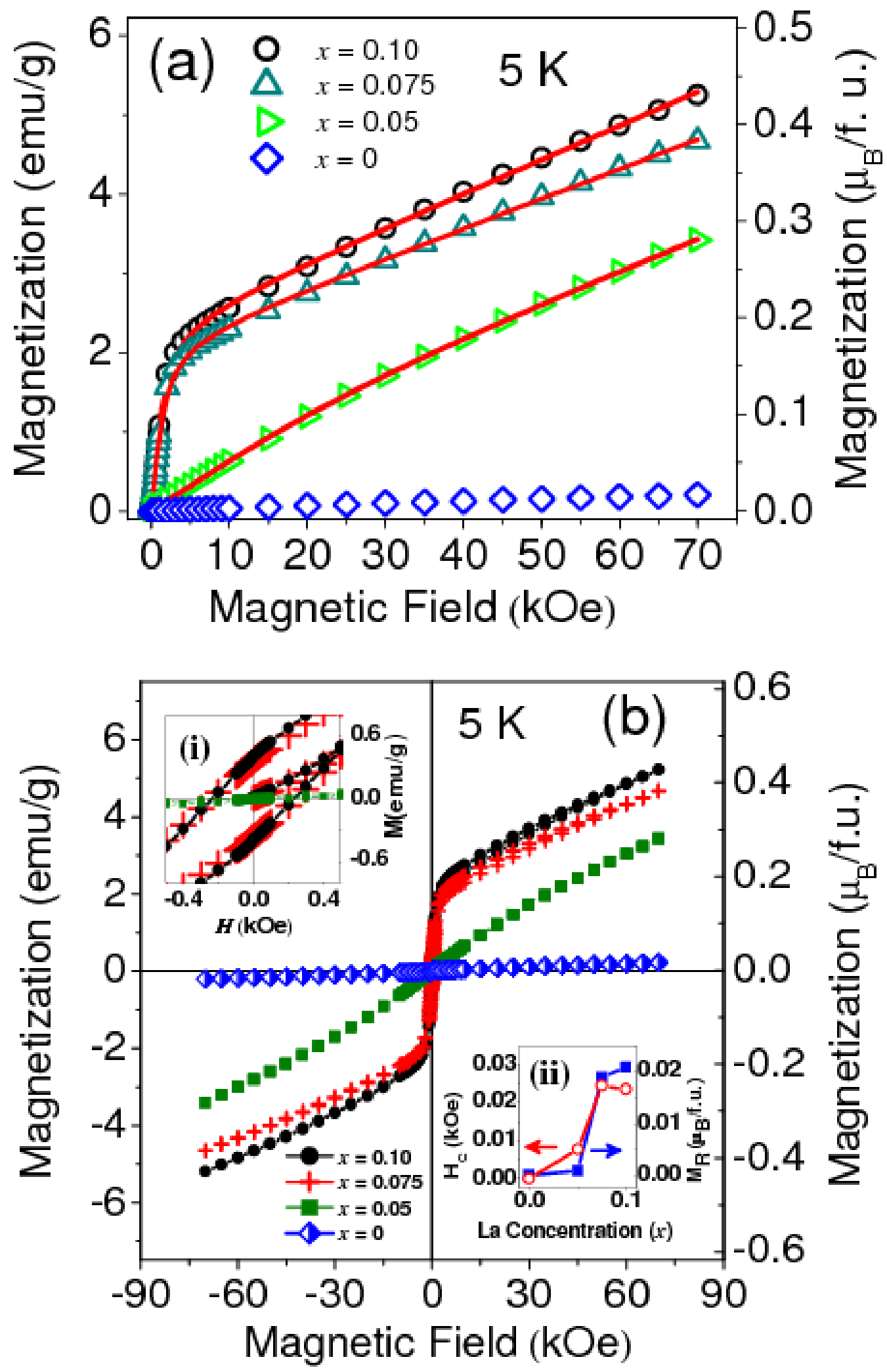

FIG. 5. (Color online) (a) The virgin magnetization curves for all four samples at 5 K. The solid curves are the fit to the observed data with Eq. (1). (b) The M vs $H$ curves over all four quadrants for $Ca_{2.5-x}La_xSr_{0.5}GaMn_2O_8$ ($x$ =0, 0.05, 0.075, and 0.10) at 5 K. Inset (i) shows an enlarge view of the M vs $H$ curves for $x$ = 0.05, 0.075 and 0.1 samples at lower fields where a clear hysteresis is seen for the $x$ = 0.075 and 0.1 samples. Inset (ii) shows the dependency of the coercive field and remanent magnetization on the La-concentration at 5 K.

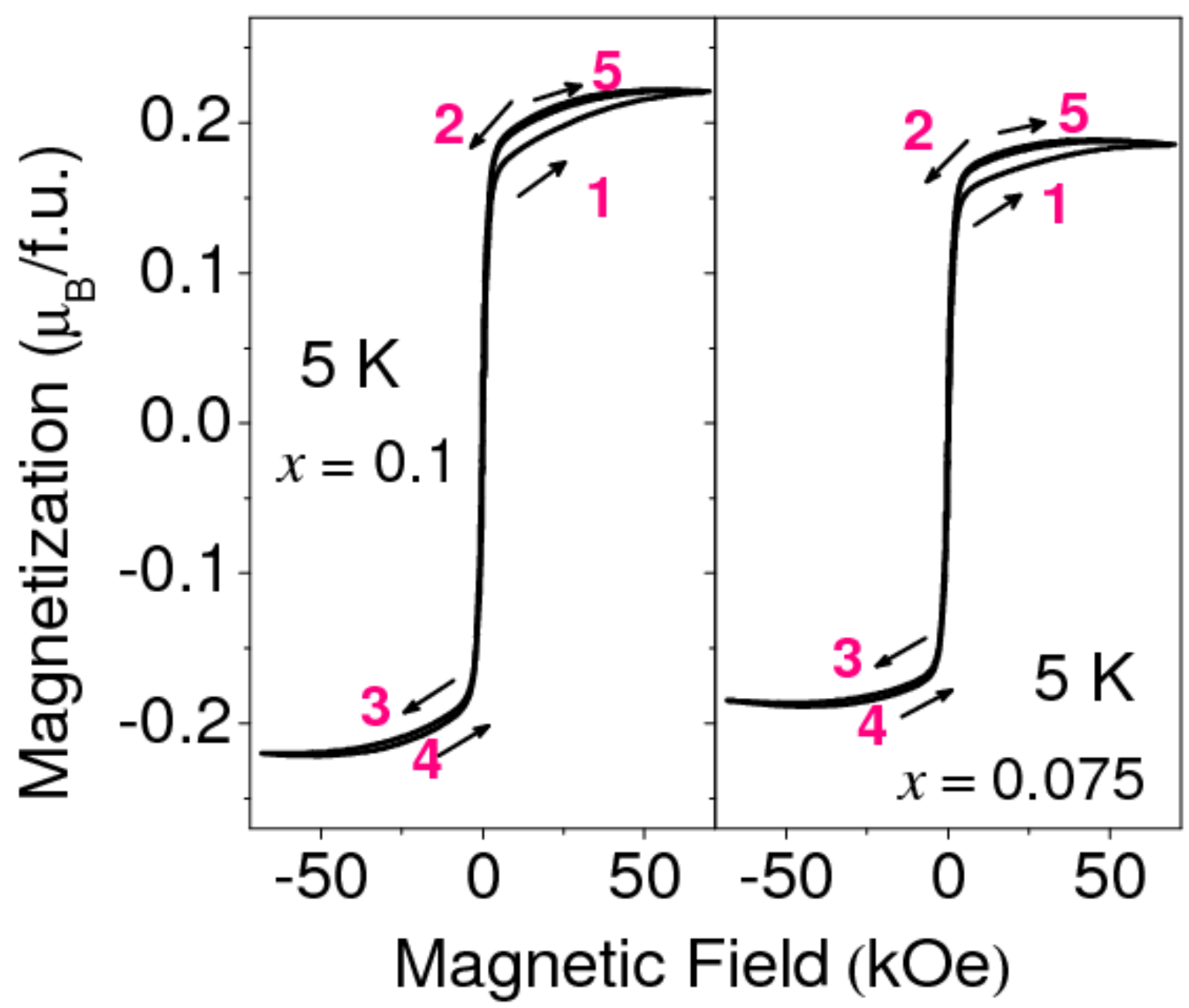

FIG. 6. (Color online) The hysteresis loop for the sample with $x$ = 0.1 and 0.075 at 5 K [after subtraction of the linear AFM contribution as per the Eq. (1)]. The numbers and the arrows indicate the sequence and the direction of the field sweep.

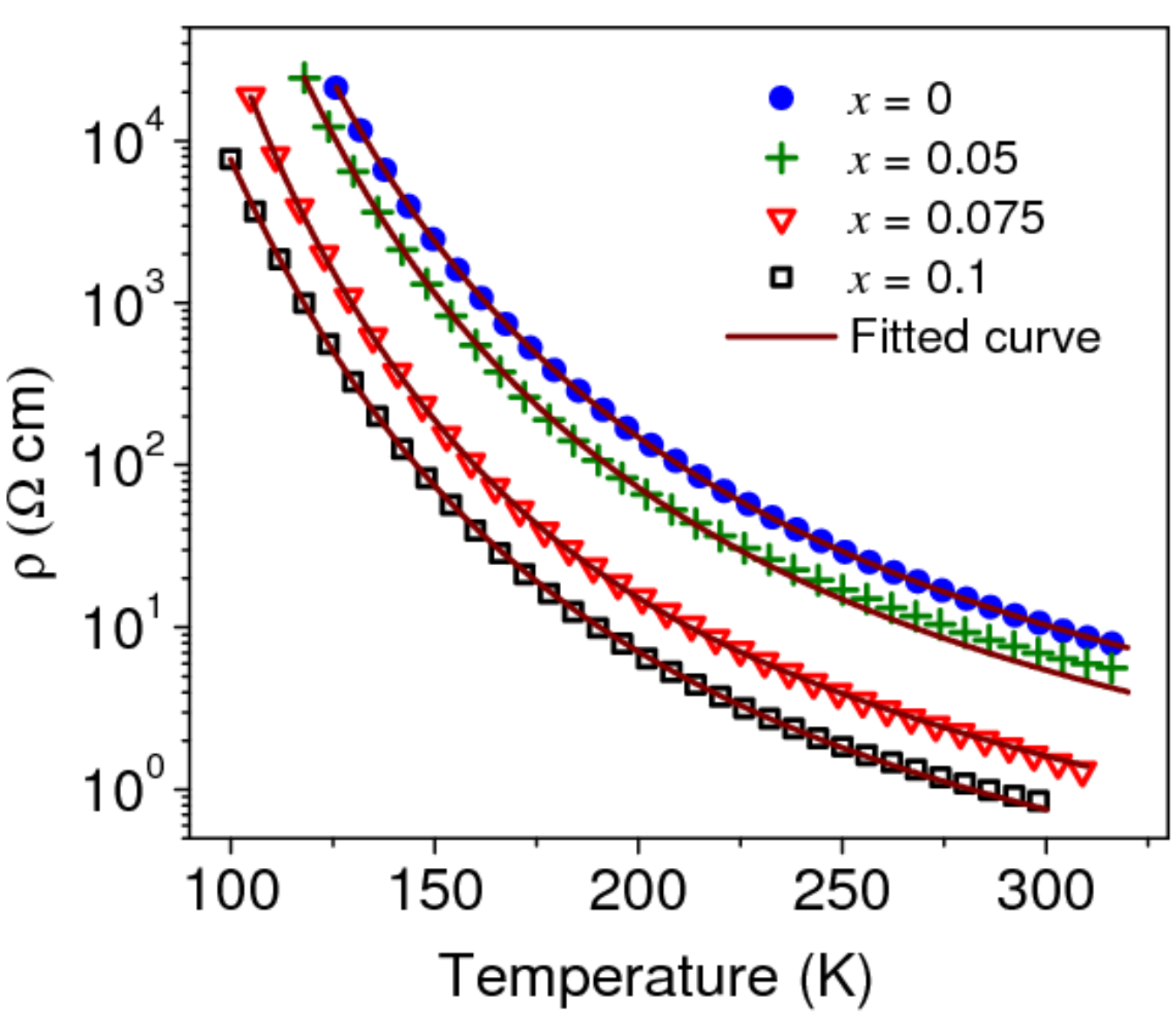

FIG. 7. (Color online) The temperature dependent electrical resistivity curves for $Ca_{2.5-x}La_xSr_{0.5}GaMn_2O_8$ ($x$ =0, 0.05, 0.075, and 0.10). Solid lines are the fitting of the observed data with the Eq. (2).

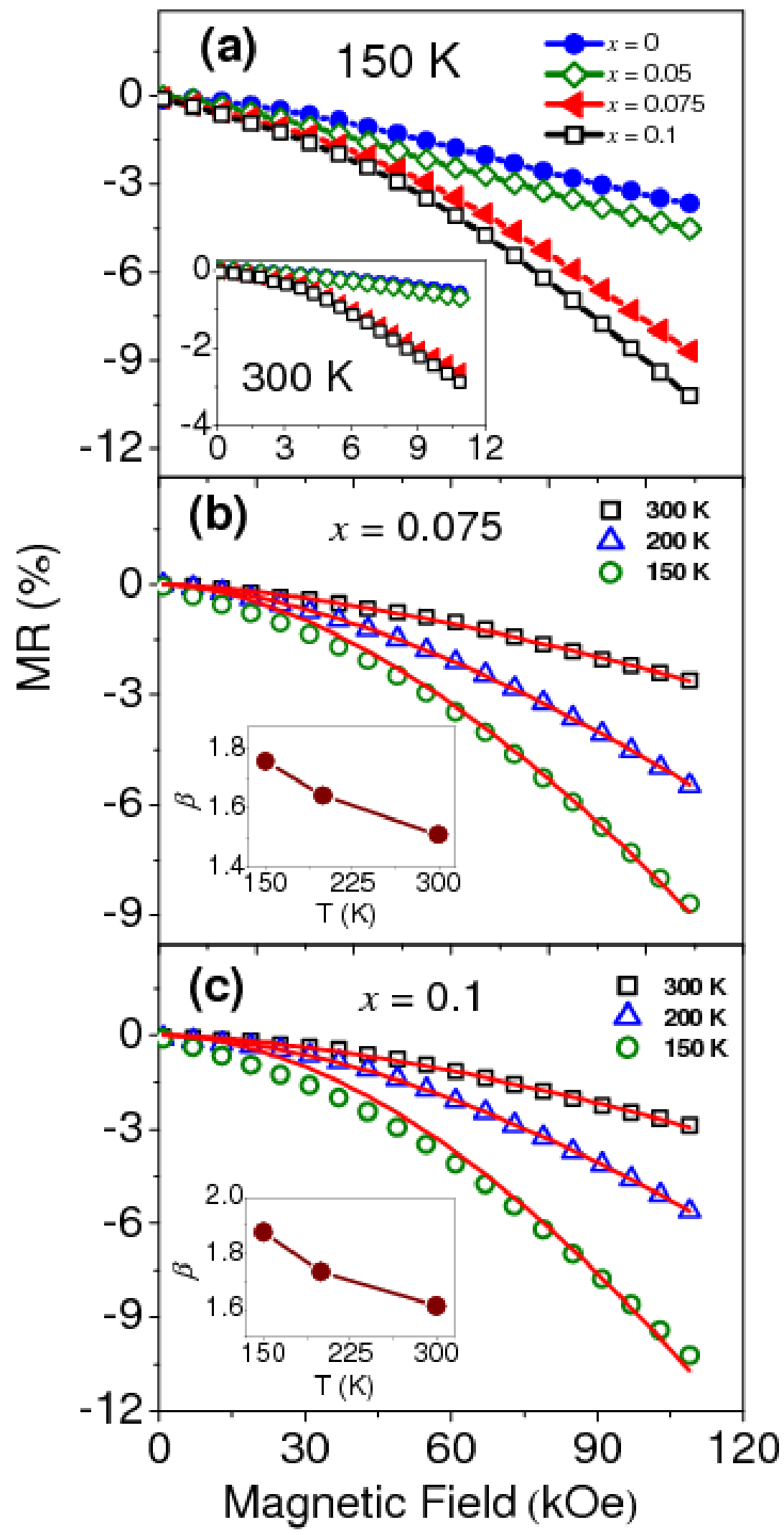

FIG. 8. (Color online) (a) The magnetoresistance $\{\mathrm{MR} = [\rho(H) - \rho(0)]/\rho(0)\}$ vs. applied magnetic field at 150 K for $Ca_{2.5-x}La_xSr_{0.5}GaMn_2O_8$ ($x$ =0, 0.05, 0.075, and 0.10). Inset shows the field dependent magnetoresistance at 300 K for all samples. (b) & (c) The dependency of the MR on magnetic field at 300, 200, and 150 K for the samples $x$ = 0.075 and 0.1, respectively. The solid lines are fit to the data according to the Eq. (8). Insets show the temperature dependence of the exponent $\beta$ for respective samples.